\documentclass[]{aa}  
\usepackage{graphicx}
\usepackage{txfonts}	
\begin{document}
   \title{Lunar dust characterization by polarimetric signature.}
	\subtitle{ I. Negative polarization branch of sphere aggregates of various porosities.}

   \author{D.T. Richard \inst{1,2}
          \and
          S.S. Davis \inst{1}
          }

   \institute{Lunar Dust Laboratory, Space Science and Astrobiology Division, NASA Ames Research Center, MS 245-3, Moffet Field CA 94035-1000, USA
        \and San Jos\'e State University Research Foundation, 210 N. Fourth St. Fourth Floor, San Jos\'e CA 95112, USA\\
        \email{[Denis.T.Richard;Sanford.S.Davis]@NASA.gov}
        }

   \date{Received / Accepted}


  \abstract
   {In support of NASA's exploration program and the return to the Moon, the polarimetric signature of dispersed individual Lunar regolith dust grains is studied to enable the characterization of the dust exopsheric environment by remote, in-situ, and standoff sensing. }
   {We explore the value of the negative polarization branch (NPB) as a signature for characterizing individual grains to determine if it can be used in the same way as for surfaces of planets and atmosphereless bodies.}
   {The linear polarization phase curve for single spheres of silicate and for aggregates of spherical silicate grains of different porosity are computed using the discrete dipole approximation (DDA) for a range of grain sizes. Features in these curves are identified and their evolution explored as a function of grain size and aggregate porosity. We focus particularly on the so-called negative polarization branch that has been historically used to characterize planetary surfaces.}
   {Calculations show that polarization phase curves for spherical grains exhibit a sharp transition over a narrow range of size parameter between two distinct regimes, one typical of Rayleigh scattering and another dominated by a large NPB. The linear polarimetric signature observed for aggregates is a composite of a) the polarization induced by individual grains composing the aggregate and b) the polarization due to the aggregate as a whole dust grain. The weight of each component varies depending on the porosity of the aggregate. An NPB similar to the one observed for atmosphereless astronomical bodies is present for different ranges of the size parameter depending on the value of the porosity. It appears as a remnant of the negative branch exhibited by the single spherical grains. The sharper, narrow negative branch that is measured for some granular surfaces in the laboratory or seen in astronomical observations is not observed here.}
   { These results suggest that the wide negative branch is due to the scattering by individual grains and single aggregates, while the narrow negative branch is more likely due to coherent backscattering or shadowing effects in bulk material. The shape and evolution of the NPB could be used to characterize spherical grains and to differentiate between aggregates with the same porosity but different sizes, but does not appear to be a practical candidate for univocally differentiating between aggregates of different porosity.}


   \maketitle


\section{Introduction}

The nature of the Lunar dust environment is an unknown that needs to be understood before long-term human exploration and settlement can become a reality. A return to the Moon will inevitably be accompanied by anthropogenic transport and dynamics.  The transcripts and debriefings of the Apollo missions demonstrate the pervasive nature of lunar dust in an exploration environment: invasion of the Lunar Module, solar panel and radiators clogging, and the possible induction of allergic reactions. Eugene Cernan, Mission Commander of Apollo 17 summarized the issue : \textit{``The dust was very difficult to work in and was a big hindrance. It obscured your vision if you let it get on your visor. It was almost like a dew or mist because it clung to everything. It got on every movable surface, it got in the suits, and when we got out of our suits in the spacecraft in between EVAs, it got in the pores of our skin and got under our fingernails. And it didn't just get on the outside parts of our nails and get them dirty but, literally, it got down between the skin and the nail. It took three months for lunar dust to grow out from under my nails. It infiltrates."} (Apollo Lunar Surface Journal (\cite{ALSJ}), Apollo 17 Journal). The composition and morphology of lunar dust have made it suspect of being toxic to humans, and its probable transport into future habitat modules and inhalation by the crew could prove to be more than a simple inconvenience. Other physical characteristics (such as high abrasiveness) are responsible for its position near the top of the list of engineering challenges. NASA has acknowledged the importance of these issues by forming the Lunar Airborne Dust Toxicity Advisory Group (LADTAG), which regroups interdisciplinary experts and has been charged with setting health standards and risk criteria for use by engineers and astronauts during lunar mission design and operation.

In addition to the transport generated by human activity, it has been suggested that dust grains in the micron and sub-micron size range can be naturally transported by various electrostatic mechanisms above the surface of the Moon to altitudes in excess of 100km (e.g., Stubbs et al. \cite{Stubbsetal2007}). Evidence of a dust exopshere includes observations of ``streamers'' and ``horizon glow'' by Apollo Astronauts (McCoy \& Criswell \cite{McCoyCriswell74}) as well as by the Surveyor landers (Rennilson and Criswell \cite{RennilsonCriswell1974}) and Apollo coronal photography (McCoy \cite{McCoy76}). The scale-height of the observed horizon glow has been estimated to be $\sim$10km (Zook \& McCoy \cite{ZookMcCoy}; Murphy \& Vondrak \cite{Murphy93}). The dusty lunar surface is electrostatically charged by the photoemission of electrons by solar ultraviolet, as well as the incident electrons and ions from the ambient plasma environment. The resulting surface electric fields drive the transport of charged lunar dust. Different mechanisms have been proposed, from the levitation of micron-size grains at $\sim$30cm above the surface, to the lofting of submicron grains to high altitudes by the dynamic fountain effect (Stubbs et al. \cite{Stubbsetal2006}). These fundamental transport mechanisms are also expected to be active on other atmosphereless solar system bodies with granular surfaces, making the results presented here relevant to other objects such as asteroids and the surfaces of airless planets and planetary satellites.

The capability to characterize the dust environment near the surface and in the atmosphere of the human habitat is critical to enable a human settlement; characterization of the grain population lofted to higher altitudes is equally important for our understanding of the fundamental transport mechanisms.  Our current work aims at developing technology to probe and characterize this environment at all altitudes by in-situ, standoff and remote sensing; the results reported in this publication are a preliminary step towards building a knowledge base to support that objective.

While early polarimetric observations of the Moon were attempted by Arago (\cite{Arago1811}, \cite{Arago1844}), the doctorate thesis of Bernard Lyot (\cite{Lyot1929}) laid the foundations for comparative studies of the polarimetric  properties of celestial bodies and terrestrial substances. Measuring the optical properties of various materials--including ash from several eruptions of mount Vesuvius--and observing the light scattered by astronomical objects he noticed that the polarization curves of the Moon, Mercury, certain areas on Mars and asteroids most resemble the curves obtained for opaque powdered materials in the laboratory. Supported by observations of integrated moonlight and regional studies, he concluded that the Moon is covered almost entirely by a powdery material closely resembling terrestrial volcanic ash. One particular feature leading to this conclusion was the presence of a negative branch on their polarization curve at small phase angles (The sign of P depending on which linear polarization component dominates the observed reflectance.) His work was followed by that of Dollfus (\cite{Dollfus}, \cite{DollfusI}, \cite{DollfusII}, \cite{DollfusIII}), who notably selected by polarimetry a fine grained, very cohesive basaltic material that reproduced satisfactorily the astronomical observations (Dollfus \cite{Dollfus0}); polarization curves from the several areas and features of the lunar surface were described in terms of specific parameters (such as the minimum and maximum of polarization, the inversion angle, the gradient at the inversion angle) and compared to laboratory measurements on meteoritic and mineralogic samples. The first samples from the lunar surface returned by Apollo 11 showed a striking resemblance with the selected material (Dollfus \cite{DollfusI}), thus validating the technique.  \{Numerous other works have generally focused on inferring the properties of particulate surfaces through the characteristics of their observed polarization phase-curves;  Geake \& Geake (\cite{GeakeandGeake1990}) and Kolokolova et al. (\cite{Kolokolovaetal1993})  studied how the polarization-phase curve of powder surfaces varies in scale and shape with grain size for sub-wavelength grains; Shkuratov et al. (\cite{Shkuratovetal2002}, \cite{Shkuratovetal2007}, \cite{Shkuratovetal2007b}) measured the scattering by particulate surfaces of various compositions and focused on the origin of the opposition effect and negative branches of the phase curve; Lasue et al. (\cite{Lasueetal2006}) and Hadamcik et al. (\cite{Hadamcik2003}, \cite{Hadamcik2006}) modeled cometary light observations by the light scattered by porous irregular aggregates of sub-micron grain sizes of silica and organics.

The origin of the negative branch, commonly associated with powdered materials and fluffy aggregates, is still debated. Lyot proposed that negative polarization is due to three mechanisms : multiple reflection, refraction in transparent particles and diffraction. While these are very broad and general, more precisely described mechanisms include shadowing or coherent backscattering in fluffy grains. The Mie solution to Maxwell's equations--Mie theory--for samples of polydispersed separated spherical grains of different refractive indices have been proposed to explain the negative polarization of cometary atmospheres. These models have not successfully been applied to planetary regolith. Shkuratov et al. (\cite{Shkuratovetal1994}) provides a detailed review of these various models and mechanisms. In addition to the wide negative branch observed by Lyot, a narrow peak (less or of the order of a degree angle wide) of negative polarization is sometimes observed in conjunction with a sharp increase in scattered intensity (opposition effect) at angles close to backscattering. It is still unclear whether both the wide and narrow negative branches have a common origin or are due to different mechanisms. Ovcharenko et al. (\cite{Ovcharenkoetal2006}) proposes--based on laboratory measurements of granular surfaces--that the wide negative polarization branch is due to individual aggregate scattering while the narrow branch is due to coherent backscattering on surface layers.  

Scanning electron photomicrographs of Apollo samples revealed the presence of a variety of grain morphology, from micrometer and sub-micrometer agglutinates with irregular and sharp edges--which origin lies in the surface bombardment by meteorites--to smoother glass droplets of volcanic origins (see McKay et al. (\cite{LunarSourceBook}) for a review of the lunar regolith properties.) But it wasn't until recently that the fine (from 100nm to 20$\mu$m) and ultrafine (20 to 100nm) particle content of the lunar regolith was measured (Greenberg et al. \cite{Greenbergetal2007}), the difficulty residing in the separation and counting technique that need to be used for sub-micron particles. Along with measuring Particle Size Distribution functions (PSD), Greenberg and collaborators also took hundreds of photomicrographs of grains; typical particles in the submicron regime--the most relevant for electrostatically transported grains--do not display any complex feature or any departure between aerodynamic and physical dimensions. They are compact ellipsoid particles with small aspect ratios. In only two photomicrographs did they observe aggregates of these compact particles. It is unknown though what proportion exists in aggregate form in the lunar environment, as the protocol to obtain PSD functions included a de-agglomeration process. Spherical grains and their aggregates are therefore a relevant approximation to electrostatically transported dust.

The present work initiates a systematic study of the polarimetric properties of individual grain models in order to characterize the lunar exospheric environment, where individual dust grains are dispersed above the surface. It is therefore different in nature than bulk or surface regolith characterization. We first describe the computation method and target models; we then present the linear polarization due to the scattering by single spheres and follow with the results for random spherical grain aggregates of various porosity. For each target type, a range of sizes is explored. We discuss the origin of the NPB and its usefulness as a dust grain characterization tool.


\section{Model and Approach}

Computations have been carried with the publicly available code DDSCAT developed by Draine \& Flatau (\cite{DraineFlatau2004}) implementing the Discrete Dipole Approximation (DDA) first proposed by Purcell \& Pennypacker (\cite{PurcellPenny1973}). The DDA approximates a continuum target by a finite array of polarizable points, therefore allowing to calculate the scattering and absorption properties of targets of arbitrary geometry. Each point acquires a dipole moment in response to the local electric field induced by the incident light wave. These dipoles interact with one another via their electric field--which is why the DDA is also referred to as the coupled dipole approximation. The method is described in more details and reviewed in Draine \& Flatau (\cite{DraineFlatau1994}). The major inputs to the code are the complex refractive index, the polarization state and frequency of the incident wave, the size, geometry and orientation of the target; outputs include the 4x4 Mueller scattering intensity matrix $S_{ij}$, as well as scattering, absorption and extinction efficiencies. The linear polarization degree is defined in terms of the Mueller matrix elements by $P=-{S_{12} / S_{11}}$.

\begin{figure}
\centering
\includegraphics[width=\columnwidth]{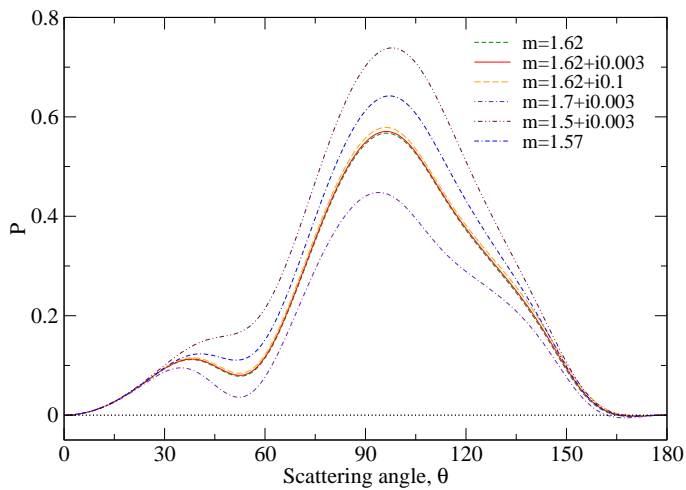}
\caption{Sensitivity of the model to variations of the complex refractive index.}
        \label{sensitivity}
\end{figure}

While the signatures of individual grains of the various shapes and compositions known for lunar material will eventually have to be determined in the future, we restrict the subject of this publication to single spherical grains and aggregates of 28 spherical grains. For aggregates, we define the porosity parameter as the percentage of void space in the target defined by the aggregate inscribed in the smallest possible box. Computations were carried out for three values of this parameter: 0.9,0.8, and 0.6. To the authors knowledge, the only measures of refractive index of lunar regolith are these of plagioclase grains from Apollo samples, of size ranging from 0.75mm to 1.55mm. These grains have a real refractive index between approximately 1.56 and 1.59 (King et al. \cite{Kingetal1971}). Because the size of these samples is  much larger than the fine grains that interest us, we chose a complex refractive index that reflects the optical properties of typical astronomical silicate dust, $m=1.62+\mathrm{i}0.003$ (Lasue \& Levasseur-Regourd 2006). The sensitivity to both the real and imaginary part of the refractive index can be seen on Fig. \ref{sensitivity}, where the polarization phase curve for an aggregate of porosity 0.6 is shown for various values of $m$. This demonstrates that practical applications of polarimetric signature characterization of the lunar dust environment will require a better knowledge than we have today of the optical properties of the fine and ultra-fine portion of the regolith. With the recent renewed interest in Lunar Science, we do not expect for this gap in our knowledge to persist for long.

   \begin{figure}
   \centering
   \includegraphics[width=\columnwidth]{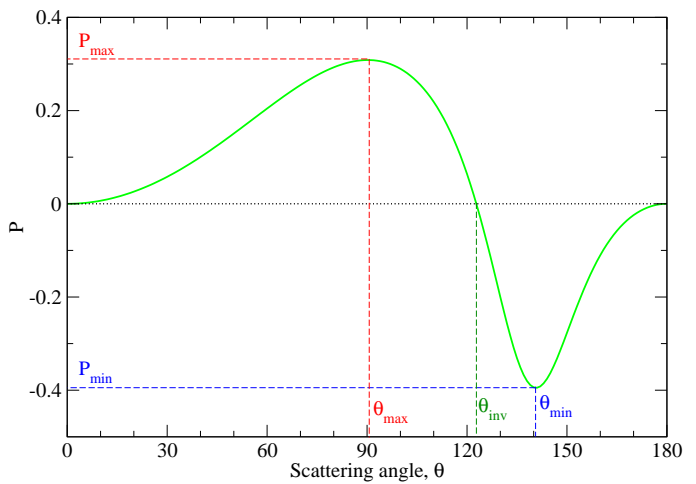}
   \caption{Notations used in this paper to describe polarization curve characteristics.}
        \label{notations}
    \end{figure}

The size parameter is varied for each kind of target; it is defined as 
$$X=ka={2 \pi a \over \lambda},$$ 
where $k$ is the wavenumber and $\lambda$ is the wavelength of the incident radiation, and $a$ is the radius of the sphere of equivalent volume. For aggregates, we define $X_{agg}$ as the size parameter of the whole aggregate, and $X_i$ as the size parameter of its constitutive individual spherical grains. The incident light is unpolarized and a reference wavelength $\lambda_0$=628.3nm is chosen for practicality ($X={2 \pi a / \lambda_0}=10 a$ with $a$ in microns). For non-spherical targets, scattering is computed and averaged over more than a thousand target orientations and is therefore representative of single scattering by a sample of separated randomly oriented grains. For each computation, the validity conditions for the DDA model are verified : $|m|kd<1$ and $d$ is smaller than any structural length scale of the target, where $d$ is the inter-dipole separation. We limit the exploration in this paper to size parameters for which a single negative branch exists in the polarization phase curve. We use the scattering angle rather than the phase angle (which is more customary in astronomy.) These two angles are supplementary; a scattering angle of 0$^\circ$ corresponds to forward scattering and 180$^\circ$ to backscattering. P$_{min}$ and P$_{max}$ refer to the maximum and minimum value of the polarization curve, and $\theta_{min}$ and $\theta_{max}$ the scattering angles for which these values are reached. The inversion angle $\theta_{inv}$ is the angle for which the polarization curve changes sign. These notations are illustrated on Fig. \ref{notations}. 

The present work is concerned with spheres and sphere aggregates of homogeneous composition; the refractive index is independent of wavelength. More complex shapes, compositions, spectral properties as well as different aspects of the polarimetric signature (beyond the characteristics of the negative polarization branch) will be considered in future publications.


\section{Polarization by single spheres}

   \begin{figure}
   \includegraphics[width=\columnwidth]{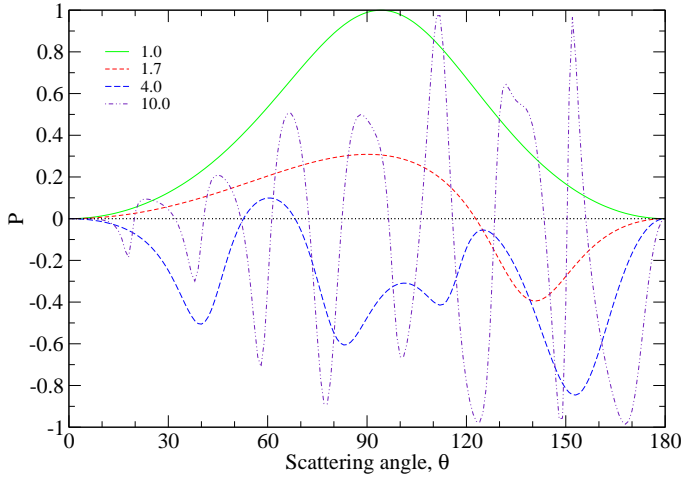}
   \caption{Linear polarization (P) of light scattered by silicate spheres of variable size parameter (X).}
   \label{SpheresModes}
    \end{figure}

The linear polarization phase curves for spherical grains of four representative size parameter are shown on Fig. \ref{SpheresModes}. For small values of $X$ (for example $X=1.0$ in this figure, corresponding at this wavelength to a grain of size $0.1 \mu m$) the phase curve is expectedly similar to that for Rayleigh scattering (for scatterers of size smaller than the wavelength): a bell-shaped curve peaking close to 90$^\circ$ scattering angle. As the size parameter increases, higher modes develop; a negative branch appears around $X=1.7$; for $X=4.0$, the curve exhibits four local minima, and for $X=10.0$ it shows 8. This behavior is classically known and is theoretically described by the Mie theory (Hulst \cite{Hulst1981}). As the behavior of the polarization curve becomes more complex, several 'negative branches' appear. In the present article, we will focus exclusively on scattering regimes where no more than one negative polarization branch (NPB) is present. For the present computations, this translates to a size parameter $X < 2.5$.

   \begin{figure}
   \includegraphics[width=\columnwidth]{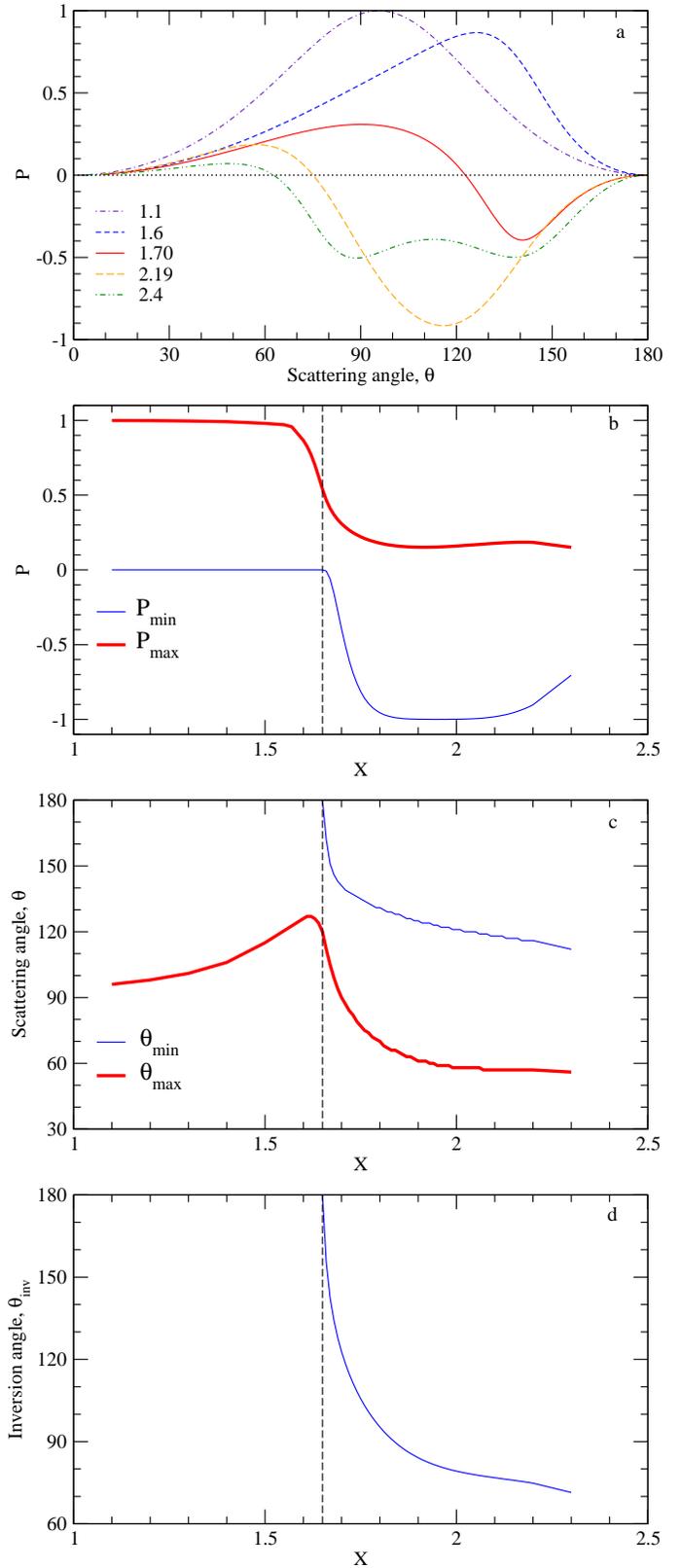}
   \caption{Linear polarization (P) phase curve of light scattered by spheres of silicate of variable size parameter ($X$); (a) P as a function of scattering angle, for five values of $X$; (b) Polarization maximum P$_{max}$ and minimum P$_{min}$ as a function of $X$; (c) Angles for which the maximum ($\theta_{max}$) and minimum ($\theta_{min}$) are reached as a function of $X$; (d) Inversion angle $\theta_{inv}$, indicating the width of the Negative Polarization Branch as a function of $X$. Vertical dashed lines mark the value of $X$ for which the NPB appears.}
   \label{Spheres}
    \end{figure}

Figure \ref{Spheres} explores the polarimetric signature and particularly the characteristics of the NPB in this size range. Figure \ref{Spheres}(a) shows the evolution of the shape of the phase curve through five representative values of $X$ as it continuously transforms from Rayleigh-typical to a curve dominated by a negative branch that is both wide and deep (with $\theta_{inv}$ close to 130$^\circ$ and P$_{min}$ close to $-1$ at $X=2$.) The Rayleigh-typical regime extends roughly until $X=1.5$; the large NPB regime is observed for $X>1.8$. This is clearly illustrated by Fig. \ref{Spheres}(b) showing the minimum and maximum of the polarization phase curve P$_{min}$ and P$_{max}$ as a function of $X$; for $X<1.5$, P$_{min}$ is equal to zero and P$_{max}$ is close to 1; for $X>1.8$ P$_{max}$ averages a much lower value of 0.15 and P$_{min}$ reaches close to $-1$.

As a precursor to the transition between the two regimes, the scattering angle for which the polarization reaches its maximum, $\theta_{max}$, first increases up to 120$^\circ$ (Fig.\ref{Spheres}(c)). The switch is then abrupt, reminiscent of a phase transition:  the polarization maximum P$_{max}$ starts to collapse for grain sizes close to $X=1.57$ (marked by a vertical dotted line on Fig.\ref{Spheres}(b) and (c)); $\theta_{max}$ then reaches a maximum (at $X=1.62$) and the curve peak starts to shift to smaller angles (Fig.\ref{Spheres}(c)); when P$_{max}$ has reached half its initial value, the negative branch appears (for $X=1.65$, marked by the dashed lined of Fig.\ref{Spheres}(b), (c) and (d)) and rapidly grows such that for $X=1.8$ it has reached down to P$_{min}$=$-0.95$ at an angle $\theta_{min}$ of 130$^\circ$ (Fig.\ref{Spheres}(c)), with an inversion angle $\theta_{inv}$ of 95$^\circ$ ( Fig.\ref{Spheres}(d)).

For values of $X>2.2$ the appearance of additional features of the curve raises the value of P$_{min}$ (see the curve for $X=2.4$ on Fig.\ref{Spheres}(a) for illustration). Polarization by larger grains shows very different signatures (as shown on Fig.\ref{SpheresModes}) and will be the object of future work.

This transition over a narrow range of size parameter $X=2\pi a / \lambda$ could provide a valuable signature for the characterization of this type of grain for size and composition. For example, by rotating scanning frequencies and finding the frequency for which this transition occurs, composition and/or size of spherical grains could be deduced.

   \begin{figure}
   \centering
   \includegraphics[width=\columnwidth]{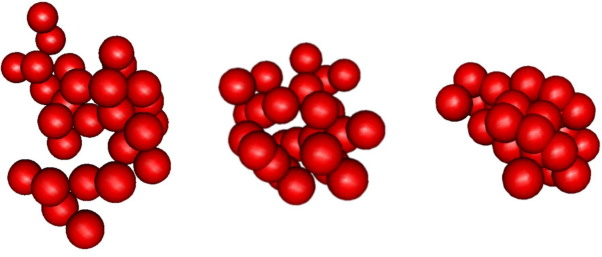}
   \caption{Prototype aggregates of porosity 0.9 (left) and 0.8 (middle) and 0.6 (right). Constitutive grains are of equal size.}
        \label{targets}
    \end{figure}


\section{Polarization by aggregates}

    Lunar regolith is partly composed of aggregates of various size, density and composition. As a first step towards a realistic description of these dust grains, we have computed the polarization phase curves for random aggregates of spherical grains. Aggregate prototypes have been created of three different porosity : 0.9, 0.8 and 0.6. These aggregates are shown in Fig. \ref{targets}. As in the single sphere case, scattering is computed for each target type for a range of size parameters.  Individual grain size scales with the aggregate. (For a given porosity, if aggregate B is twice smaller than aggregate A, the grains in B are also twice smaller than in A.) This choice is voluntary and does not reflect a limitation of the model or the DDA code. In this light, various values of $X$ for a given aggregate porosity can be seen as experiments where
 
\begin{itemize}
        \item either the incident wavelength is constant and illuminates different aggregates composed of individual grains of variable size;
        \item or the physical size of the aggregate is constant and the incident wavelength is variable.
\end{itemize}

   \begin{figure}
   \centering
        \includegraphics[width=3.25in]{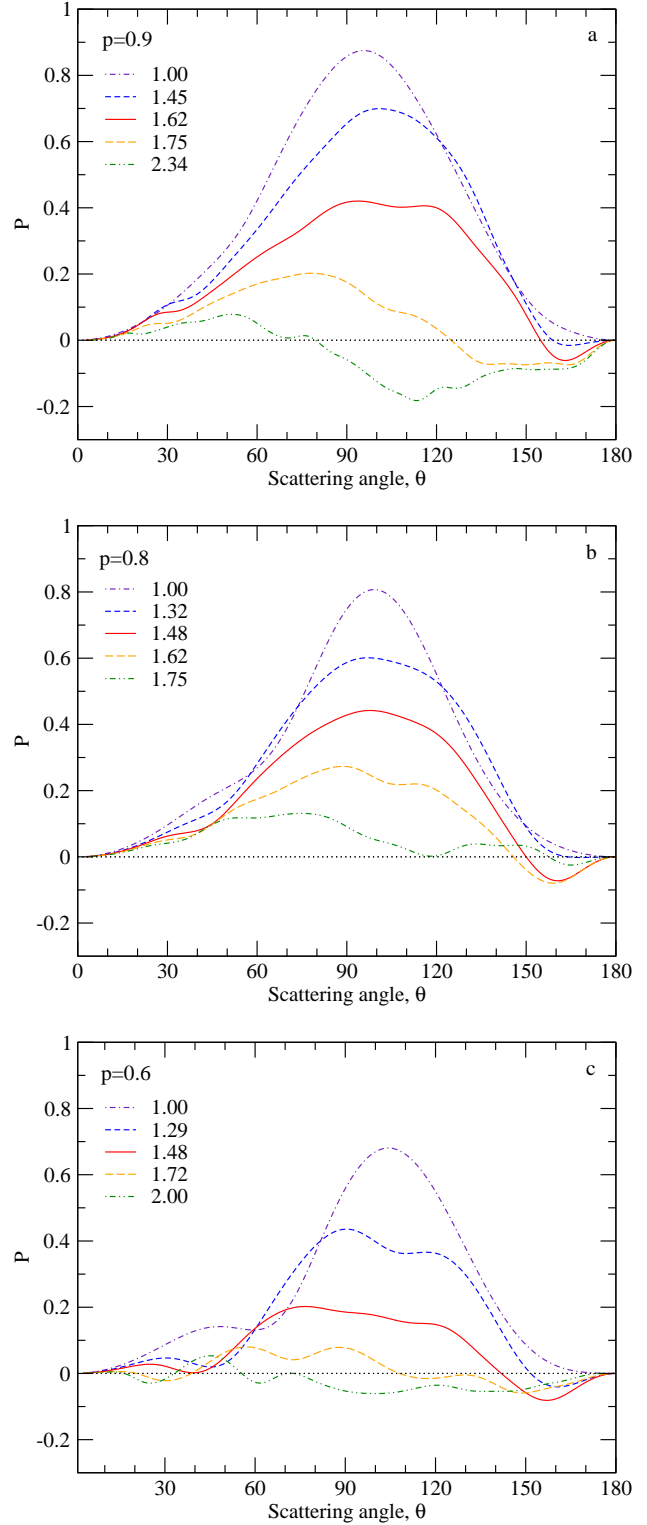}
   \caption{Linear polarization phase curve for aggregates of silicate spheres of porosity (a) p=0.9, (b) p=0.8 and (c) p=0.6. For each porosity five values of $X_i$ are shown.}
   \label{AggregatesCurves}
    \end{figure}

The linear polarization phase curves for the three porosity values are shown in Fig. \ref{AggregatesCurves}, for several values of the constitutive grain size parameter $X_i$. As the size parameter increases as well as when the porosity decreases, the general trend is (1) the appearance of higher modes, leading to an increased irregularity and (2) a decrease in the amplitude of the polarization degree curve. For $X_i=1$ and for a porosity of 0.9 the curve very much resembles Rayleigh scattering; As the porosity decreases, the maximum shifts to larger scattering angles, a secondary peak appears between 40$^\circ$ and 50$^\circ$ and the curve  becomes bi-modal. Further evolution can be seen on the curve for aggregate porosity of 0.6, at $X_i=1.29$, where a third mode is apparent (Fig. \ref{AggregatesCurves}(c)).  

The regime dominated by a large and deep NPB observed for spherical grains is replaced for aggregates by shallower, irregular polarization curves; the width of the NPBs remains comparable to the spherical case. The evolution of the polarization phase curve resembles the evolution experienced over a much larger range of size parameters by the spherical targets (Fig.\ref{SpheresModes}) but the aggregate curves develop higher modes faster. The amplitude of polarization is lower than in the spherical case. This is probably a result of multiple scattering between individual constitutive grains. Depolarization of an initially polarized incident wave due to multiple scattering is a known phenomenon (see for example Mishchenko et al. (\cite{Mishchenkoetal2006})); as an unpolarized wave is polarized by single scattering on the first constitutive grain it encounters, subsequent scattering occurring before its emergence from the aggregate will result in some depolarization.

   \begin{figure}
   \centering
   \includegraphics[width=\columnwidth]{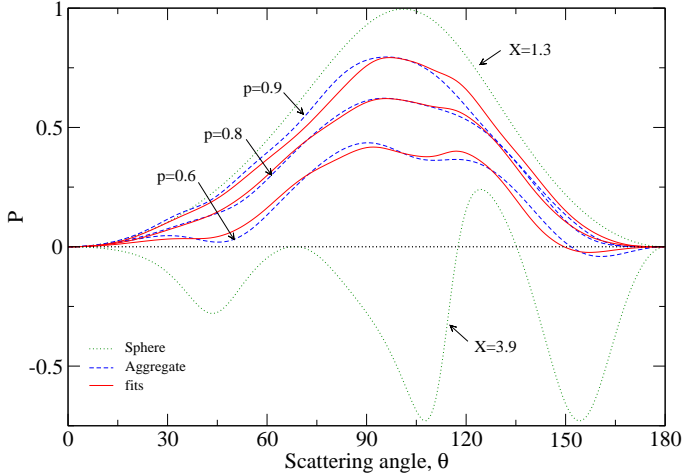}
   \caption{Linear polarization phase curves for aggregates of three different porosity, of global size parameter $X_agg=3.9$ and individual grain size parameter $X_i=1.3$ (dashed lines), and for spheres of size parameters $X=1.3$ and $X=3.9$ (dotted lines). Solid lines are best fit assuming the superposition of single scattering by a target of size parameter $X=3.9$ and multiple-scattering up to the ninth order by targets of $X=1.3$.}
    \label{interpole}
    \end{figure}

This behavior suggests that the polarization induced by these aggregates can be described as the result of the superposition of the scattering due to (1) each individual constitutive grain and (2) the scattering due to the larger sized aggregate. This is illustrated by Fig.\ref{interpole} showing the polarization curve for aggregates of size parameter $X_{agg}=3.9$ whose individual constitutive spherical grains have a size parameter $X_i=1.3$ (dashed lines). The curves for spheres of size $X=3.9$ and $X=1.3$ are also shown (dotted lines). For each target, the solid line curve is the best fit of the superposition of (1) single scattering by a spherical target of $X=3.9$, (2) single scattering by spherical targets of size $X=1.3$, (3) multiple-scattering up to the ninth order by spherical targets of size $X=1.3$, as described by: 

$$
{\cal P}(\theta)= A_0 P(\theta,X=3.9)+\sum_{i=1}^n A_i P^n(\theta,X=1.3),
$$

where $\cal P$ is the fitting function, $P(\theta,X)$ is the polarization curve for a silicate sphere of size parameter $X$ and $A_0,...,A_n$ are fitting parameters; $n$=9 for this figure. Despite the simplicity of this formulation, the resulting curves exhibit features similar to the ones seen for the actual aggregate. It is evident from these curves that the more compact is the aggregate, the more does the polarization curve reflect the features of scattering by a spherical target of its size parameter. Conversely and expectedly, the more porous the aggregate, the closer is its curve to scattering and multiple scattering by spheres the size of each individual grain.

   \begin{figure}
   \centering
   \includegraphics[width=3.25in]{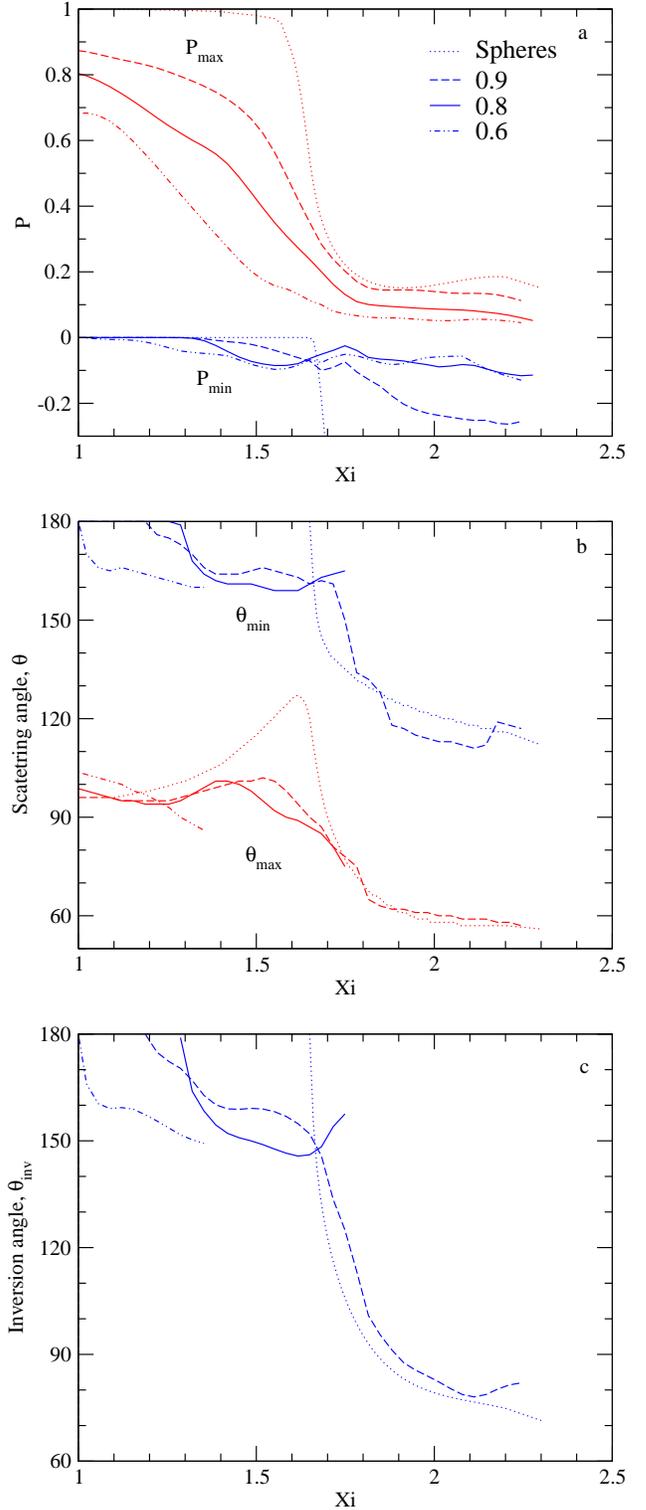}
   \caption{ Characteristics of the linear polarization phase curves of aggregates: (b) P$_{min}$ and P$_{max}$, (c) $\theta_{min}$ and $\theta_{max}$, (d) $\theta_{inv}$, as a function of X. In (b) and (c) curves are plotted only for the range of $X$ for which a single NPB exist.}
   \label{AggregatesProps}
    \end{figure}

As with the spherical case, the behavior described above is quantified by the parameters shown in Fig. \ref{AggregatesProps}. The evolution of P$_{min}$ with $X_i$ illustrates the shallowness of the negative polarization branches compared to the spherical case; Fig. \ref{AggregatesProps}(a) shows that the evolution of P$_{max}$ with $X$ is more gradual as the porosity decreases. Curves representing $\theta_{inv}$ in Fig. \ref{AggregatesProps}(c) show that there is no regular monotonic behavior identifiable for the appearance of the negative branch as a function of the porosity ; when $p$ varies from 0.9 to 0.8 and 0.6, the size parameter for which the negative branch appears goes from 1.19 to 1.28 and 1.0. The same conclusion holds for $\theta_{min}$ and $\theta_{max}$ shown in Fig. \ref{AggregatesProps}(b). It is nevertheless still clear from these figures that the higher the porosity, the closer the curves are to the case of spherical individual spheres.


\section{Conclusions}

We have compared the degree of linear polarization of light scattered by spheres of silicate and aggregates having three different values of porosity as a function of grain size. We have focused on the range of size parameter for which a negative branch exists at high scattering angle (low phase angle). For spherical grains, the regime transition over a narrow range of size parameter at the onset of the appearance of the NPB makes it a good candidate for single grain characterization; conducting multi-spectral observations of the lunar exosphere and identifying the wavelength for which a sharp change in polarization occurs would yield information about size and composition of grains. For aggregates it has been shown that, for high porosity, the behavior is close to that of individual spheres. As the porosity decreases, the curve becomes closer to that of a larger sphere of size parameter similar to that of the whole aggregate. While the NPB can provide a useful signature to differentiate between aggregates of the same porosity but of different sizes, it would be difficult to differentiate univocally between aggregates of different porosity, thus requiring complementary measurements (e.g. phase function shape) to further constrain the geometry.

Our results suggest that the wide NPB in these aggregates is a remnant of that observed in individual constitutive grains. Ovcharenko et al. (\cite{Ovcharenkoetal2006}) proposes that the wide NPB of planetary surfaces finds its origin in the individual aggregates, while the narrow NPB--which is unseen in our calculations of single grains--is the result of coherent backscattering between aggregates on a granular surface. Combining our results with these from Shkuratov et al. (\cite{Shkuratovetal2007}) and Ovcharenko et al. (\cite{Ovcharenkoetal2006}), we observe that the wide NPB, which is largest for individual spherical grains, decreases in amplitude for aggregate and decreases further for granular surface, essentially originates from the scattering (single and multiple) by individual grains constituting the aggregates. As the polarization for individual aggregates evolves from a Rayleigh typical bell-curve to a marked NPB to finally a rather flat, low polarization degree curve, it seems probable that the negative branch that is always observed for surfaces such as the Moon's is caused by the averaging over a distribution of these phase curves. Aggregates of small size parameter contribute to the positive maximum; large size parameters average to negligible contributions; intermediary size parameters contribute to the negative branch.

We have shown the possibilities and limitations of using the NPB as a polarimetric signature for the lunar dust environment in the simple case of spherical silicate grains and aggregates of spheres. Further calculations and analysis are currently being pursued to develop a thorough model of scattering properties of dispersed lunar regolith taking into account the various shapes, structures and agglomeration states found in its granular matter, as well as the range of sizes and the varying elemental composition.


\begin{acknowledgements}
     The authors would like to thank Dr. J.Cuzzi for his support, as well as Dr. T.J.Stubbs and another anonymous referee for their helpful comments. Computations supporting this publication have been conducted using the freely available DDSCAT software developed by B.T. Draine and P.J. Flatau. Figures have been created with the Grace plotting software, the R software environment for statistical computing and graphics, and the rgl package. This research has made use of NASA's Astrophysics Data System. 
\end{acknowledgements}


\end{document}